# HARD X-RAY IMAGING OF INDIVIDUAL SPECTRAL COMPONENTS IN SOLAR FLARES

AMIR CASPI[1], ALBERT Y. SHIH[2], JAMES M. MCTIERNAN[3], AND SÄM KRUCKER[3,4]
[1]Southwest Research Institute, Boulder, CO 80302, USA
[2]Solar Physics Laboratory, NASA Goddard Space Flight Center, Greenbelt, MD 20771, USA
[3]Space Sciences Laboratory University of California, Berkeley, CA 94720, USA
[4]Institute of 4D Technologies, School of Engineering, University of Applied Sciences and Arts Northwestern Switzerland, 5210 Windisch, Switzerland

## ABSTRACT

We present a new analytical technique, combining *Reuven Ramaty High Energy Solar Spectroscopic Imager* (*RHESSI*) high-resolution imaging and spectroscopic observations, to visualize solar flare emission as a function of spectral component (e.g., isothermal temperature) rather than energy. This computationally inexpensive technique is applicable to all spatially-invariant spectral forms and is useful for visualizing spectroscopically-determined individual sources and placing them in context, e.g., comparing multiple isothermal sources with nonthermal emission locations. For example, while extreme ultraviolet images can usually be closely identified with narrow temperature ranges, due to the emission being primarily from spectral lines of specific ion species, X-ray images are dominated by continuum emission and therefore have a broad temperature response, making it difficult to identify sources of specific temperatures regardless of the energy band of the image. We combine *RHESSI* calibrated X-ray visibilities with spatially-integrated spectral models including multiple isothermal components to effectively isolate the individual thermal sources from the combined emission and image them separately. We apply this technique to the 2002 July 23 X4.8 event studied in prior works, and image for the first time the super-hot and cooler thermal sources independently. The super-hot source is farther from the footpoints and more elongated throughout the impulsive phase, consistent with an *in situ* heating mechanism for the super-hot plasma.

## 1. INTRODUCTION

Solar flares explosively release large amounts of magnetic energy, a significant fraction of which goes into transient heating of coronal plasma to temperatures up to tens of mega-Kelvin (MK). Numerous observations have shown that hot, ~5–25 MK plasma is ubiquitous in flares of all scales (Fletcher *et al.* 2011); results from the *Reuven Ramaty High Energy Solar Spectroscopic Imager* (*RHESSI*; Lin *et al.* 2002) have also shown that so-called "super-hot" ($\gtrsim$30 MK) plasmas are common in intense, X-class flares (Caspi *et al.* 2014a) and that they are distinct from the ~5–25 MK plasma (Caspi & Lin 2010, hereafter CL10).

*RHESSI* provides high spectral and spatial resolution for X-ray observations down to ~3 keV, enabling precise measurements of the thermal continuum from plasmas with temperatures $\gtrsim$10 MK; it is most sensitive to the hottest plasmas, and thus is ideal for studying super-hot flares. While the ~5–25 MK plasma is commonly accepted to result from "evaporation" of chromospheric material heated by collisions of flare-accelerated electrons during the impulsive phase (Holman *et al.* 2011), evidence suggests that the super-hot plasma is heated directly in the corona — potentially within the acceleration region — via a fundamentally different physical process (e.g., Masuda 1994; Masuda *et al.* 1998; CL10; Longcope & Guidoni 2011; Caspi *et al.* 2014a). However, it has not been possible to directly visualize the super-hot plasma separately from the cooler, chromospherically-evaporated plasma. Extreme ultraviolet (EUV) imagers observe in spectral lines whose temperature sensitivity is weak or nonexistent above ~20 MK, while X-ray images of continuum emission contain contributions from multiple temperatures, making it difficult to identify specific thermal components.

CL10 combined *RHESSI* image data in multiple energy bands with spatially-integrated spectral models to derive centroid positions of the super-hot and cooler isothermal components for the 2002 July 23 X4.8 flare; the centroid separation was statistically significant. Here, we describe a powerful mathematical technique to directly manipulate *RHESSI* X-ray visibility image data and apply it to the same event to derive spatially-resolved image maps of the two thermal components, for the first time revealing their morphologies rather than just their centroids. This spatial information provides new insight into the physical origins of these two sources. Throughout the impulsive phase of the flare, the super-hot source is farther from the nonthermal footpoints, and more elongated, supportive of an *in situ* heating mechanism for the super-hot plasma, as suggested by earlier works. While motivated here by thermal sources, this visibility manipulation technique is applicable to arbitrary spectral distributions, including nonthermal power laws (Brown 1971) and kappa distributions (Oka *et al.* 2013).

## 2. METHOD DETAILS

EUV imagers, such as the Atmospheric Imaging Assembly (AIA; Lemen *et al.* 2012) onboard the *Solar Dynamics Observatory* (*SDO*; Pesnell *et al.* 2012), typically operate in wavelength bands dominated by spectral lines from ions with relatively narrow ranges of formation temperature (e.g., Mazzotta *et al.* 1998), allowing a quasi-one-to-one mapping between energy/wavelength and temperature (the passbands typically contain lines from multiple ions at different formation temperatures, but the average temperature response is usually dominated by a relatively narrow range, e.g., Boerner *et al.* 2014). In contrast, because solar X-rays are predominantly continuum emission and/or from relatively hot ($\gtrsim$10 MK) spectral lines with broad formation temperature ranges, X-ray images at a single energy/wavelength can nonetheless contain contributions from a wide range of plasma temperatures. Thus, *RHESSI* images made over arbitrarily narrow energy bands (Hurford *et al.* 2002) still retain broad temperature sensitivity, and it is not straightforward to determine whether emission at a given location is from thermal plasma at a particular temperature (see Figure 1). A similar problem applies to other spectral forms, e.g., distinguishing between thermal and nonthermal sources. Even if sources are spatially separated, an image in a single energy band does not allow for a determination of the source spectrum, since source intensity depends on multiple parameters (e.g., both temperature and emission measure), requiring measurements in multiple energy bands to fully specify the parameter space. Imaging spectroscopy — analyzing the flux within selected regions of numerous images in multiple energy bands — enables recovery of spectra within specific regions, but does not allow for direct visualization of individual spectral sources, e.g., in terms of temperature. Here, we present a method to accomplish this.

*RHESSI* is a Fourier imager, encoding the spatial frequencies of sources as temporal frequencies in measured lightcurves (Hurford *et al.* 2002), analogous to an interferometer. The imaging data can be expressed as calibrated visibilities[1], complex numbers with amplitude and phase encoding the measured Fourier components (2D spatial frequencies) in $(u,v)$ space (Hurford *et al.* 2005), analogous to visibilities in radio interferometry (Monnier 2003). Visibilities can be summed linearly, allowing direct manipulation of the image data that would otherwise not be well-defined.

The visibility $V$ at spatial-frequency point $(u,v)$, for photon energy $E$, can be written as the product of the integrated source flux $F$ at that energy, and a "relative" visibility $v$ that encodes the flux-normalized spatial parameters (morphology), as

$$V(u,v,E) = F(E)\, v(u,v,E) \quad (1)$$

Consider two spatially-distinct sources on the Sun, $\mathcal{S}_A$ and $\mathcal{S}_B$, each having a spectral shape that does not vary in space (e.g., an isothermal source). Because visibilities add linearly, the total observed visibilities are the sums of the individual source visibilities, as (henceforth omitting the $u,v$-dependence)

$$V_{tot}(E) = V_A(E) + V_B(E)$$
$$\Rightarrow F_{tot}(E)\, v_{tot}(E) = F_A(E)\, v_A + F_B(E)\, v_B \quad (2)$$

The $F_i$ represent the individual source intensities at energy $E$, while the $v_i$ represent the individual source morphologies, which are independent of energy by construction. Note that $v_{tot}$, the morphology of the combined source, is *not* independent of energy, since the combined spectrum is not spatially uniform; this is evident from Equation (2) and is visualized in Figure 1.

To construct images of the individual sources, we must recover $v_i$, but Equation (2) is not invertible since the $F_i$ are not independently measurable. However, from the spatially-integrated spectral model (see Figure 2), we know that $\mathcal{S}_A$ contributes some fraction $f$ of the total emission at energy $E$, and consequently, $\mathcal{S}_B$ contributes $(1-f)$ of the flux, hence

$$F_A(E) = f(E)\, F_{tot}(E); \quad F_B(E) = (1-f(E))\, F_{tot}(E) \quad (3)$$

Thus, combining Equations (2) and (3),

$$F_{tot}(E)\, v_{tot}(E) = [f(E)\, F_{tot}(E)]\, v_A + [(1-f(E))\, F_{tot}(E)]\, v_B$$
$$\Rightarrow v_{tot}(E) = f(E)\, v_A + (1-f(E))\, v_B \quad (4)$$

Because $v_{tot}$ is directly observed, and $f$ is known from spectroscopy, we now have sufficient information for inversion; observations at two energies, $E_1$ and $E_2$, uniquely determine the relative visibilities $v_A$ and $v_B$ of the individual sources, as

$$v_A = [(1-f(E_1))\, v_{tot}(E_2) - (1-f(E_2))\, v_{tot}(E_1)] / [f(E_2) - f(E_1)]$$
$$v_B = [f(E_2)\, v_{tot}(E_1) - f(E_1)\, v_{tot}(E_2)] / [f(E_2) - f(E_1)] \quad (5)$$

This inversion is well-defined so long as the denominator, $\Delta f$, is not close to zero. (If the energies can be chosen such that $f(E_2) \approx 1$ and $f(E_1) \approx 0$, the inversion becomes trivial: $v_A \approx v_{tot}(E_2)$, and $v_B \approx v_{tot}(E_1)$; in practice, $\Delta f$ need only be sufficiently non-zero for the quotient to not diverge.) For isothermal sources, because the spectrum decreases exponentially, with $e$-folding of ~2 keV for temperatures of ~20–40 MK (CL10), this can be ensured by requiring that $E_2 - E_1 \gtrsim e^{2/(T_2-T_1)}$, with $E_i$ and $T_i$ expressed in keV.

Equation (5) is generalizable to $N$ sources, requiring observations at $N$ different energies, as

$$\begin{bmatrix} v_{tot}(E_1) \\ v_{tot}(E_2) \\ \vdots \\ v_{tot}(E_N) \end{bmatrix} = \begin{bmatrix} f_{\mathcal{S}_1}(E_1) & f_{\mathcal{S}_2}(E_1) & \cdots & 1-\sum_{i=1}^{N-1} f_{\mathcal{S}_i}(E_1) \\ f_{\mathcal{S}_1}(E_2) & f_{\mathcal{S}_2}(E_2) & \cdots & 1-\sum_{i=1}^{N-1} f_{\mathcal{S}_i}(E_2) \\ \vdots & \vdots & \ddots & \vdots \\ f_{\mathcal{S}_1}(E_N) & f_{\mathcal{S}_2}(E_N) & \cdots & 1-\sum_{i=1}^{N-1} f_{\mathcal{S}_i}(E_N) \end{bmatrix} \cdot \begin{bmatrix} v_{\mathcal{S}_1} \\ v_{\mathcal{S}_2} \\ \vdots \\ v_{\mathcal{S}_N} \end{bmatrix} \quad (6)$$

More compactly,

$$\mathbf{v_{tot}} = \mathbf{F} \cdot \mathbf{v_{\mathcal{S}}} \quad \Rightarrow \quad \mathbf{v_{\mathcal{S}}} = \mathbf{F}^{-1} \cdot \mathbf{v_{tot}} \quad (7)$$

where the individual source visibilities $\mathbf{v_{\mathcal{S}}}$ can therefore be recovered as long as the fractional contribution matrix $\mathbf{F}$ is invertible. Such generalization allows this visualization technique to be applied to multiple spectral components, e.g., to more than two isothermals or to a binned or parametrized multi-component emission measure distribution (e.g., Caspi *et al.* 2014b).

Because *RHESSI* records individual photons containing both spectral and spatial information (Hurford *et al.* 2002; Smith *et al.* 2002), the images and spectra are derived from the same data, and hence so are the relative combined visibilities $v_{tot}$ and the fractional contributions $f_{\mathcal{S}_i}$. The individual source visibilities $v_i$ are thus recovered completely self-consistently. This method can be considered complementary to imaging spectroscopy — instead of using images to determine spatially-resolved spectra, we manipulate image data to visualize already-determined spatially-integrated spectral models. It is also complementary to electron visibility maps (Piana *et al.* 2007; Massone & Piana 2013) since, despite reconstructing the underlying electron populations, an electron map at a single energy can still contain contributions from multiple populations (e.g., two distinct nonthermal components). Additionally, the photon-to-electron inversion necessarily assumes a single emission mechanism and is therefore not able to simultaneously treat an ensemble of different physical processes as our technique can.

This technique is mathematically simple, computationally inexpensive, and fully automatable, as long as the spectral model has already been computed at each time interval to provide the fractional contribution $f$ (or $\mathbf{F}$).

## 3. APPLICATION AND ANALYSIS

To highlight the power of this technique, we applied it to the 2002 July 23 X4.8 flare (see Lin *et al.* 2003, and references therein). CL10 showed that two spectrally-distinct sources exist throughout this event: a "super-hot," ~21–44 MK source, and a "hot," ~13–24 MK source. While they derived a significant centroid separation, their analytical approach could not reveal the detailed source morphologies.

We used the precisely-calibrated spectroscopic results from CL10 to determine the $f(E)$ values (Figure 2). To maximize $\Delta f$ in the denominator of Equation (5), we chose 1-keV-wide energy bins around $E_1 \approx 6.83$ keV (the Fe line complex; Phillips 2004) and $E_2 = 17.5$ keV (arbitrarily chosen to be dominated by the super-hot component but well below the nonthermal regime), coinciding with two of the energy bands selected by CL10 for their centroid analysis (see their Figure 2 inset, and Caspi 2010 for full details). For each spectroscopy time interval, visibilities $V_{tot}$ were obtained from the `hsi_visibility` object within the *RHESSI SolarSoft*[2] IDL software package and normalized by their intensities to derive the relative visibilities $v_{tot}$; we used all detectors ex-cept 1 (insufficient modulation from the sources), 2 and 7 (noisy and thus not suitable for use below ~20 keV). The routine `hsi_vis_combine.pro` was used to linearly combine the $v_{tot}$ following Equation (5) to determine the individual source visibilities ($v_{T_1}$, $v_{T_2}$) for the two isothermal components. We then used the UV_SMOOTH algorithm[3] (Massone *et al.* 2009) to reconstruct images from the visibilities, both in energy (combined sources) and in temperature (individual sources).

---

[1] http://sprg.ssl.berkeley.edu/~tohban/wiki/index.php/RHESSI_Visibilities

[2] http://www.lmsal.com/solarsoft/

Figure 3 shows the results of this analysis for a single time interval at the flare soft X-ray (SXR) peak. A movie spanning ~32 min of the impulsive phase and early decay is included online; selected frames are assembled in Figure 4. The morphologies derived through this novel technique reveal the detail only hinted at by the CL10 centroids. The super-hot source is not only farther from the hard X-ray (HXR) footpoints compared to the cooler source, it is also significantly elongated away from them, which in this projected geometry is either higher into the corona or farther along the loop arcade (or both). This morphology supports a coronal *in situ* heating mechanism for the super-hot component during or just following the reconnection process, whence hotter plasma would appear in higher, newly-reconnected loops while cooler plasma formed through chromospheric "evaporation" would reside in older, lower loops. CL10 proposed this interpretation for the flare onset, but could only speculate for later times. Our novel technique shows, for the first time, that the elongation and source separation persist during the impulsive phase and early decay, indicating that this mechanism likely operates throughout the flare.

To evaluate the accuracy and robustness of our thermal image reconstruction technique against noise and variable source separation, we utilize *RHESSI*'s simulation software[4] (also within *SolarSoft*) to model two idealized sources at known locations with different (known) thermal spectra and generate lists of detected photons in a Monte Carlo fashion. From these simulated photons, we can generate visibilities and reconstruct images in fixed energy bands (Figure 5, upper left), as well as images of the two individual thermal sources using the same technique as above (Figure 5, upper right). For relevance, the spectral and spatial parameters of the two simulated sources are chosen from the best-fit parameters from the peak of the 2002 July 23 flare (Figures 2 and 3), with the simplification that the sources are represented using elliptical Gaussians. We can then compare the reconstructed source morphologies, locations, and centroid separation against the true (input) values as the latter are varied. For the single simulated example in Figure 5 (top), the cooler source reconstruction is biased noticeably toward the super-hot source, as is apparent both from the peak location and the 50% contour.

The intensities of the simulated sources are matched to the counting statistics actually observed in this flare. We determine the mean and standard deviation of the reconstructed spatial parameters by analyzing an ensemble of simulations. The reconstructed source locations are determined using `hsi_vis_fwdfit.pro`, which fits an elliptical Gaussian source to the visibilities. For a range of input source separations (4–16″), with 25 runs at each point, the reconstructed source centroid separation is generally ~0.5–1.5″ smaller than the true input value, with the largest deviation observed at a separation of ~12″ (Figure 5, lower left). This discrepancy is dominated by inaccuracies in reconstructing the cooler source (Figure 5, lower right), which is not unexpected as this source has poorer counting statistics and hence larger relative uncertainties.

Therefore, we find that this technique can reconstruct source centroid separations with a mean inaccuracy of ≲10%, with variance due to counting statistics. The detailed source morphology is more challenging to reconstruct, although the magnitudes of the contour discrepancies are difficult to quantify. Nonetheless, the gross morphology appears reasonably reconstructed in all cases. A more in-depth characterization of this technique will involve simulating a much broader space of source parameters. For actual observations, the inaccuracy in source locations can be approximated by the uncertainties reported by `hsi_vis_fwdfit.pro`; alternatively, the actual observed visibilities can be perturbed by their uncertainties and a Monte Carlo analysis performed on the reconstructed sources, similar to our analysis with simulated sources, which would also help quantify the uncertainties of the individual source morphologies.

## 4. DISCUSSION

The ease of application of this technique enables this kind of study – distinguishing between overlapping thermal sources – to be performed on many flares. For example, the 14 super-hot flares from the survey of Caspi *et al.* (2014a) could be analyzed in the same way as the 2002 July 23 event, to study the morphological properties of the super-hot and cooler thermal components to determine whether the relationship discovered here holds generally. Furthermore, because this technique can be applied using a wide range of spectral distributions, a number of other applications are enabled, such as distinguishing between overlapping thermal and nonthermal sources (e.g., during the pre-impulsive phase of 2002 July 23; CL10) or separating the sources of electron and ion gamma-ray emission (e.g., Shih *et al.* 2009).

The spatial diagnostic provided by this technique allows for direct comparison between the identified sources and other contextual information. For example, we note an intriguing feature observed just after the SXR peak, around 00:32:30 UT (Figure 4, panel 3) — an apparent HXR source appears in the corona, co-spatial with the derived super-hot source. It contains ~20–30% of the total 60–100 keV flux at this time, as bright as the footpoints. Although the *RHESSI* imaging software does not currently account for pulse pile-up, preliminary analysis suggests that this source represents real HXR emission in the corona. While a chromospheric origin cannot be completely discounted due to projection, there is no corresponding loop footpoint in EUV or Hα (cf. Lin *et al.* 2003), and this feature is distinct from the "middle" footpoint identified by Emslie *et al.* (2003), supporting a coronal origin.

All of this information combined provides a deeper physical insight into the origins of super-hot plasma. That the HXR source is co-spatial with the elongated super-hot source, and *not* with the cooler source, suggests that the super-hot source is in the same loop as the accelerated particles, either within or downstream from the reconnection region. The nonthermal emission indicates that the loop must be relatively newly reconnected, supporting the *in situ* heating mechanisms proposed by CL10 or by Longcope & Guidoni (2011) for the super-hot plasma. Oka *et al.* (2013) suggested a kappa distribution in the acceleration region, which could simultaneously explain co-spatial super-hot and nonthermal emission with a single spectral model, although here the centroids and morphologies of the two sources are not identical, complicating this interpretation. We will investigate this source and its connection with the super-hot plasma, including accounting for the effects of pulse pile-up, in a future paper.

Although here applied to *RHESSI* visibilities, our technique can be adapted to work with any calibrated imaging data. In particular,

---

[3] N.B. UV_SMOOTH requires disabling the addition of conjugate visibilities in the `hsi_visibility` object.

[4] http://hesperia.gsfc.nasa.gov/rhessi2/home/software/simulation-software/

AIA observes in six EUV passbands sensitive to coronal temperatures, although the temperature response of many of the passbands includes multiple peaks (Boerner *et al.* 2012). Methods exist to approximately isolate these peaks (e.g., Warren *et al.* 2012; Del Zanna 2013), but they rely on ad hoc, empirical corrections. Our technique can be directly applied to AIA images, using spectra from the EUV Variability Experiment (EVE; Woods *et al.* 2012) to determine a thermal emission model (e.g., Warren *et al.* 2013; Caspi *et al.* 2014b) from which the individual temperature contributions to the various AIA passbands can be determined, enabling the same kind of linear combination of wavelength-space images to be performed to recover temperature-space, model-dependent, derived images of the individual thermal (but not necessarily *isothermal*) components. Although the imaging and spectral data would come from different instruments, AIA image fluxes are calibrated against EVE irradiances (Boerner *et al.* 2014), preserving self-consistency. Temperature maps reconstructed with this method can be compared to pixel-by-pixel differential emission measure calculations using AIA data alone (e.g., Aschwanden & Boerner 2011; Hannah & Kontar 2012). Our technique could be similarly applied to full-Sun SXR images from the X-ray Telescope (XRT; Golub *et al.* 2007) onboard *Hinode*, with spectral information obtained from the upcoming *Miniature X-ray Solar Spectrometer* CubeSat (Caspi *et al.* 2015; Mason *et al.* 2015).

This work was supported by NASA contract NAS5-98033. AC and JMM were also supported by NASA grants NNX08AJ18G and NNX12AH48G. We thank G. Hurford and R. Schwartz for many helpful discussions.

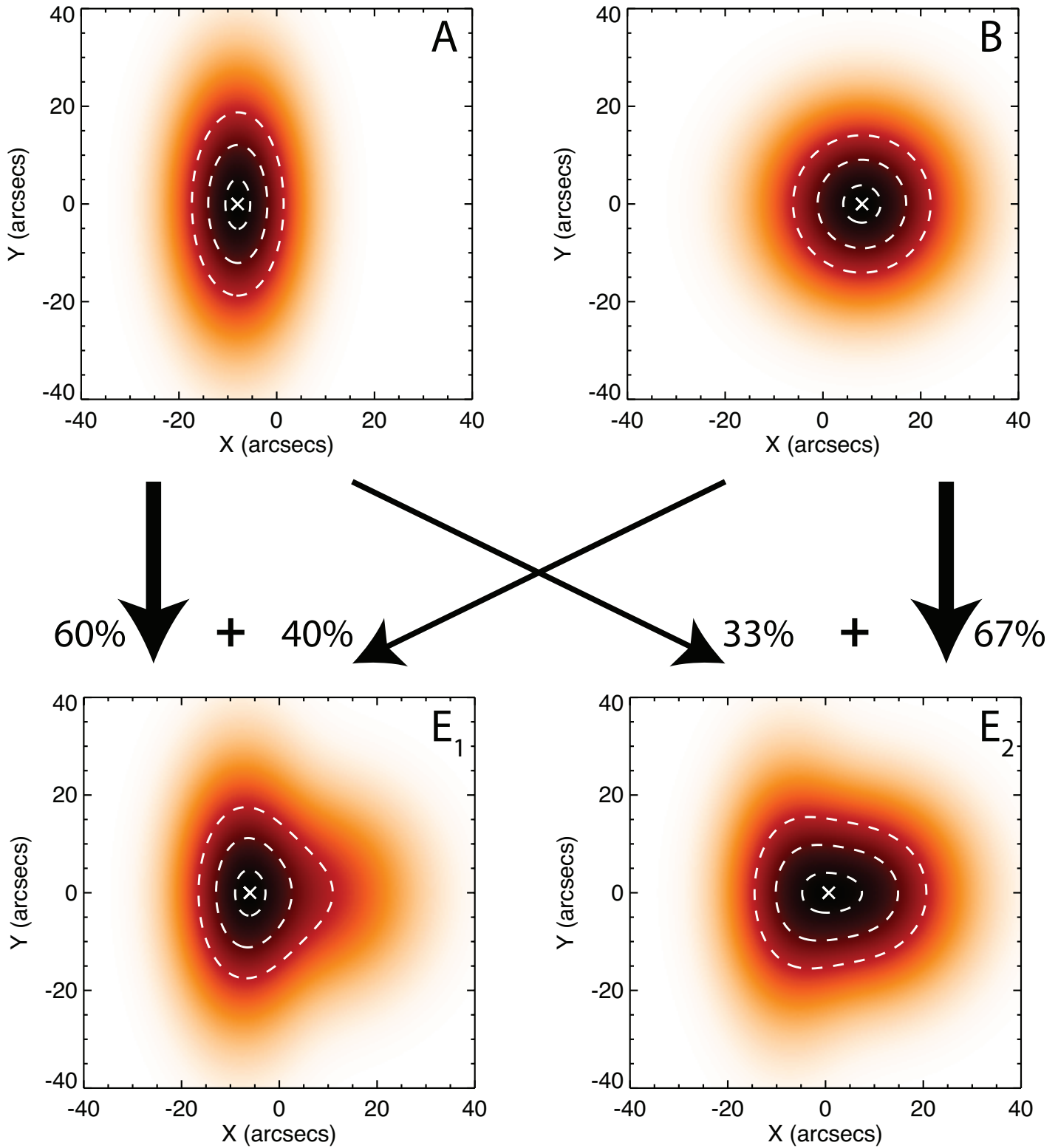

**Figure 1** – Schematic representation of sources adding in *RHESSI* images. Sources *A* and *B* (*top*) have different spectra $\mathcal{S}_A$ and $\mathcal{S}_B$ whose shapes do not vary in space (e.g., isothermal sources); although their *intensities* vary with energy, their *morphologies* (i.e., shapes of the isobrightness contours) do not. If the fractional contributions of the two sources to the total intensity vary with energy, the combined source (*bottom*) will have varying morphology at different energies.

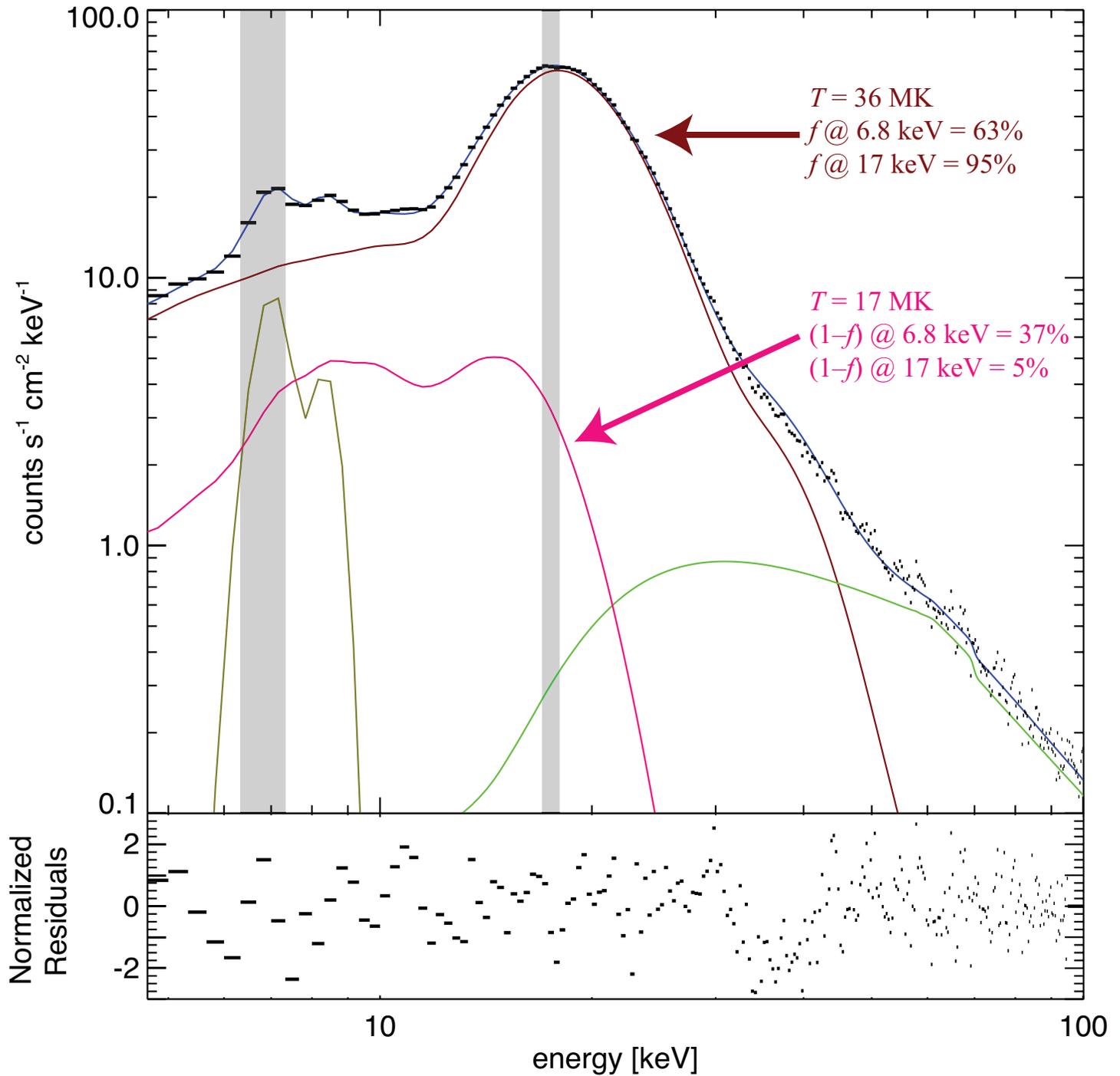

**Figure 2** – *RHESSI* count flux spectrum, with model fit (2 isothermals – *brick, magenta*; Fe/Fe–Ni lines – *mustard*; nonthermal – *green*), during the SXR peak of the 2002 July 23 X4.8 event. From the model, we can compute the fractional contribution of each isothermal component, to linearly combine the total visibilities and reconstruct the individual source visibilities. The shaded spectral regions (6.33–7.33 and 17–18 keV) are used for the reconstruction shown in Figures 3 and 4.

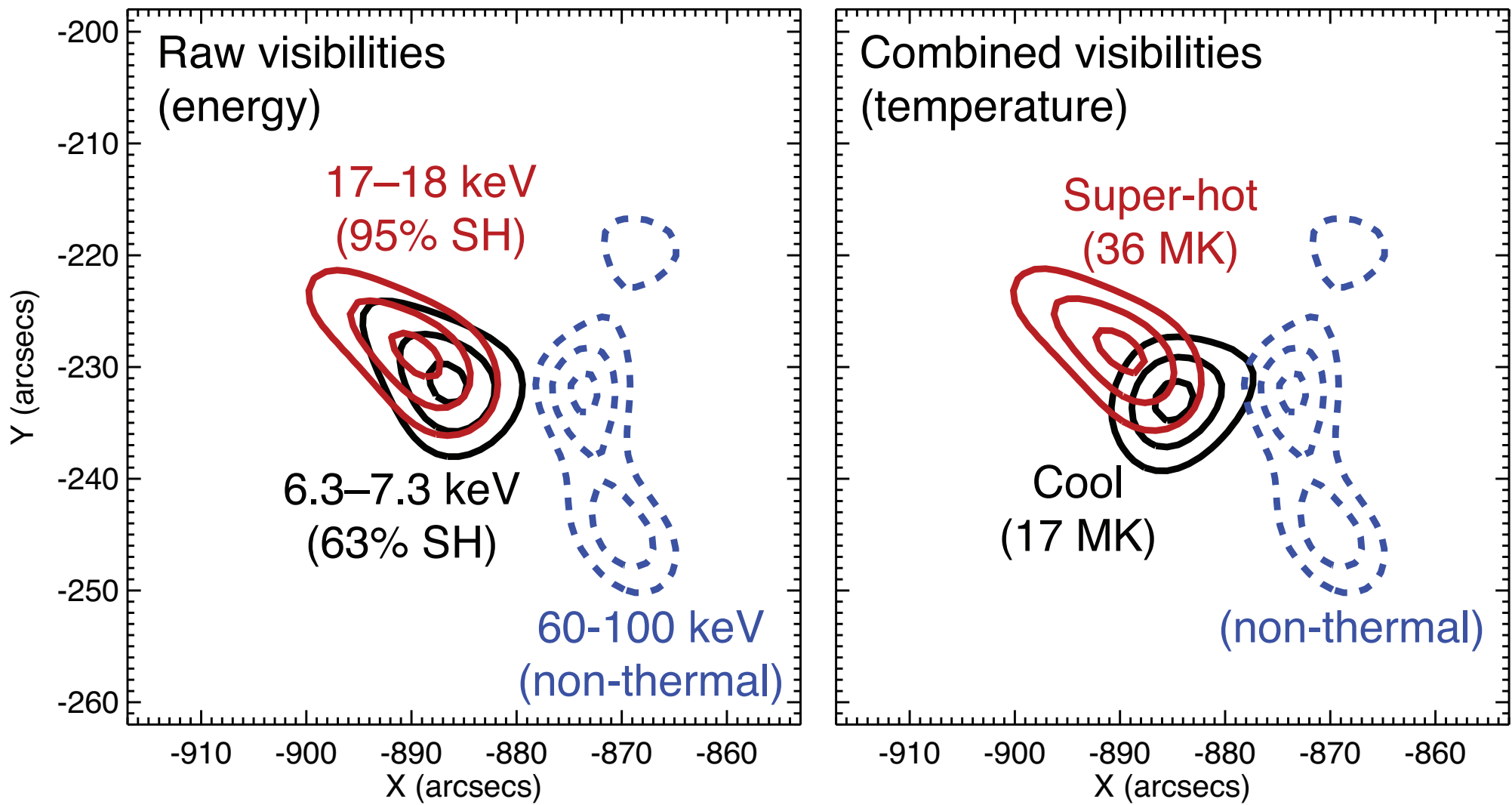

**Figure 3** – Visibilities at two different energies (*left; red, black*) are combined with the fractional contribution of the super-hot source to the total flux at each energy to yield energy-independent reconstructed images of the two thermal sources (*right*), per Equation (5). (The 60–100 keV nonthermal emission, *blue dashed*, is shown for reference.) Contours are at 50%, 75%, and 95% of peak intensity for each source.

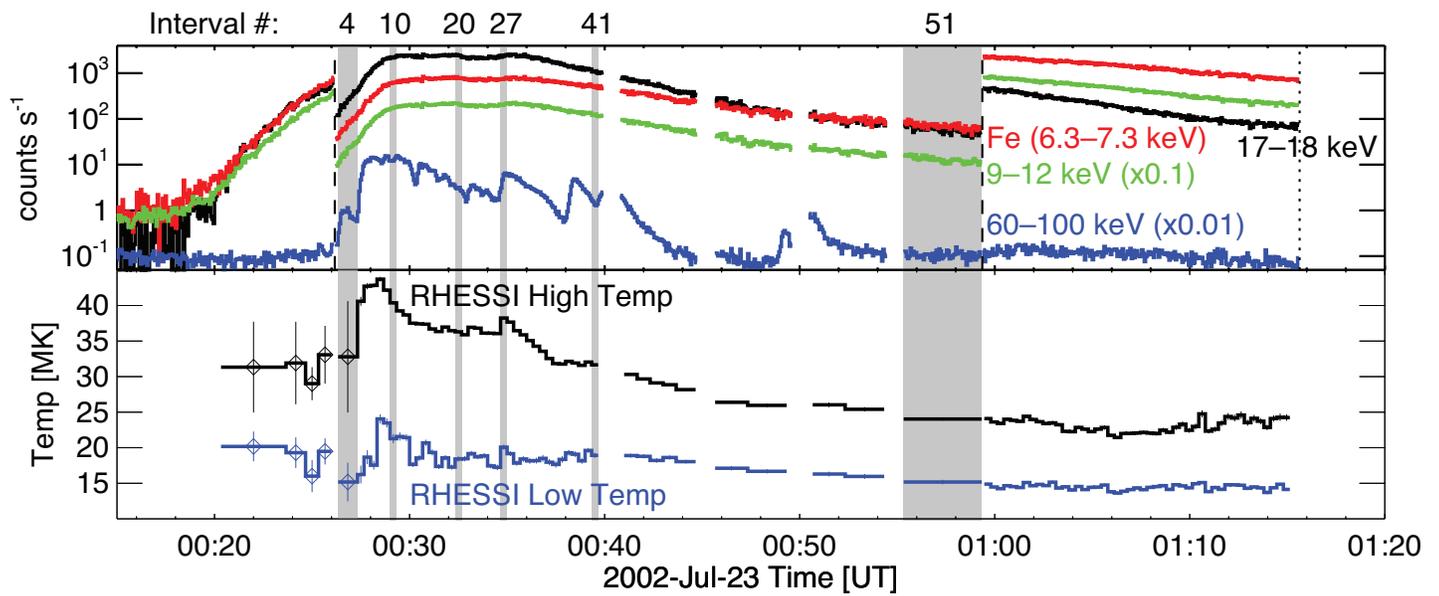
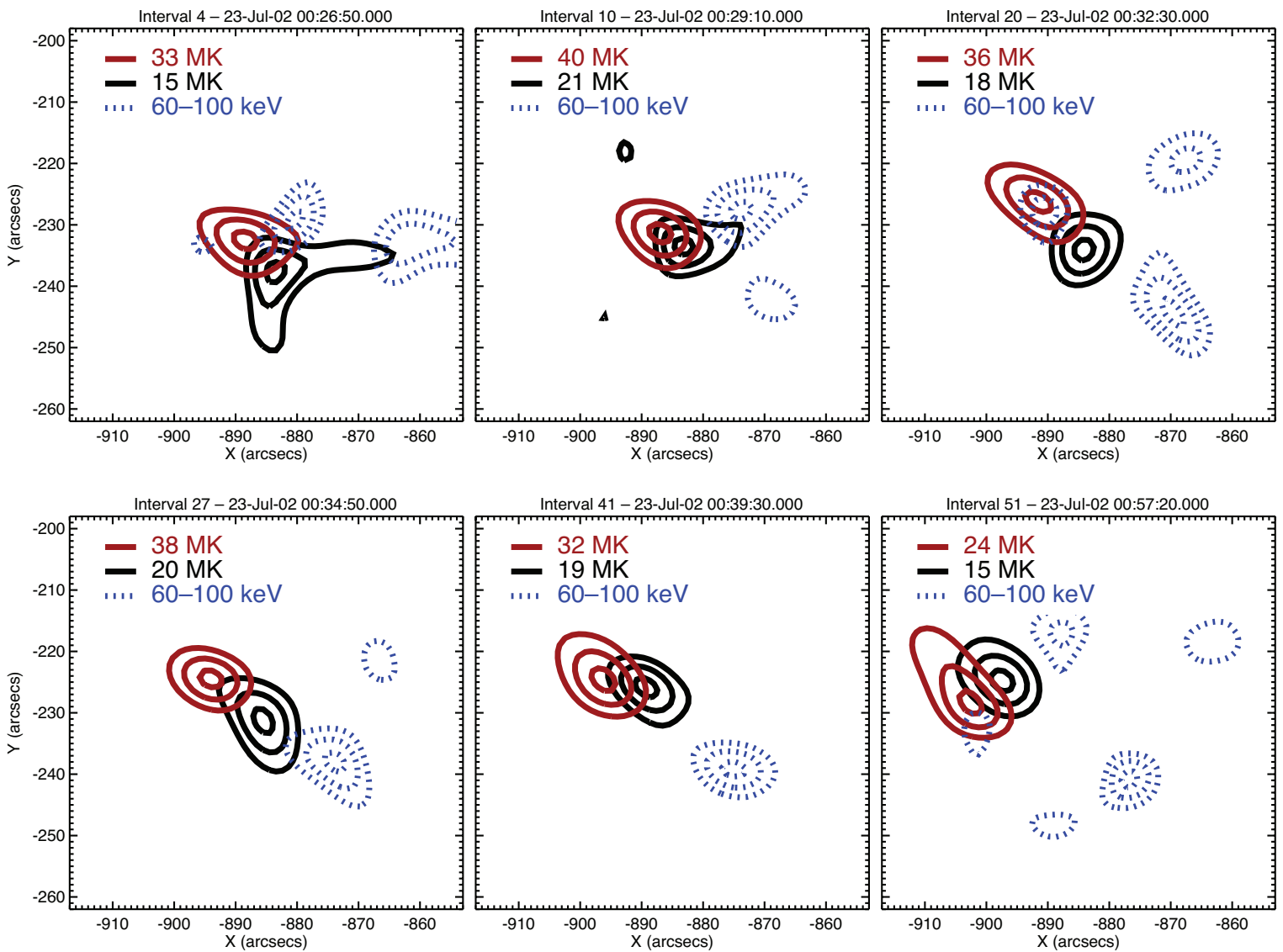

**Figure 4** – Selected frames from the online movie showing the reconstructed thermal sources and their evolution over time; the 60–100 keV nonthermal footpoints are shown for reference (in panel 6, these contours are noisy due to poor statistics). The super-hot source is always farther from the footpoints, and more elongated, compared to the cooler source. An intriguing HXR, likely nonthermal source is visible in the corona, cospatial with the super-hot source, in panel 3. Contour levels are the same as for Figure 3.

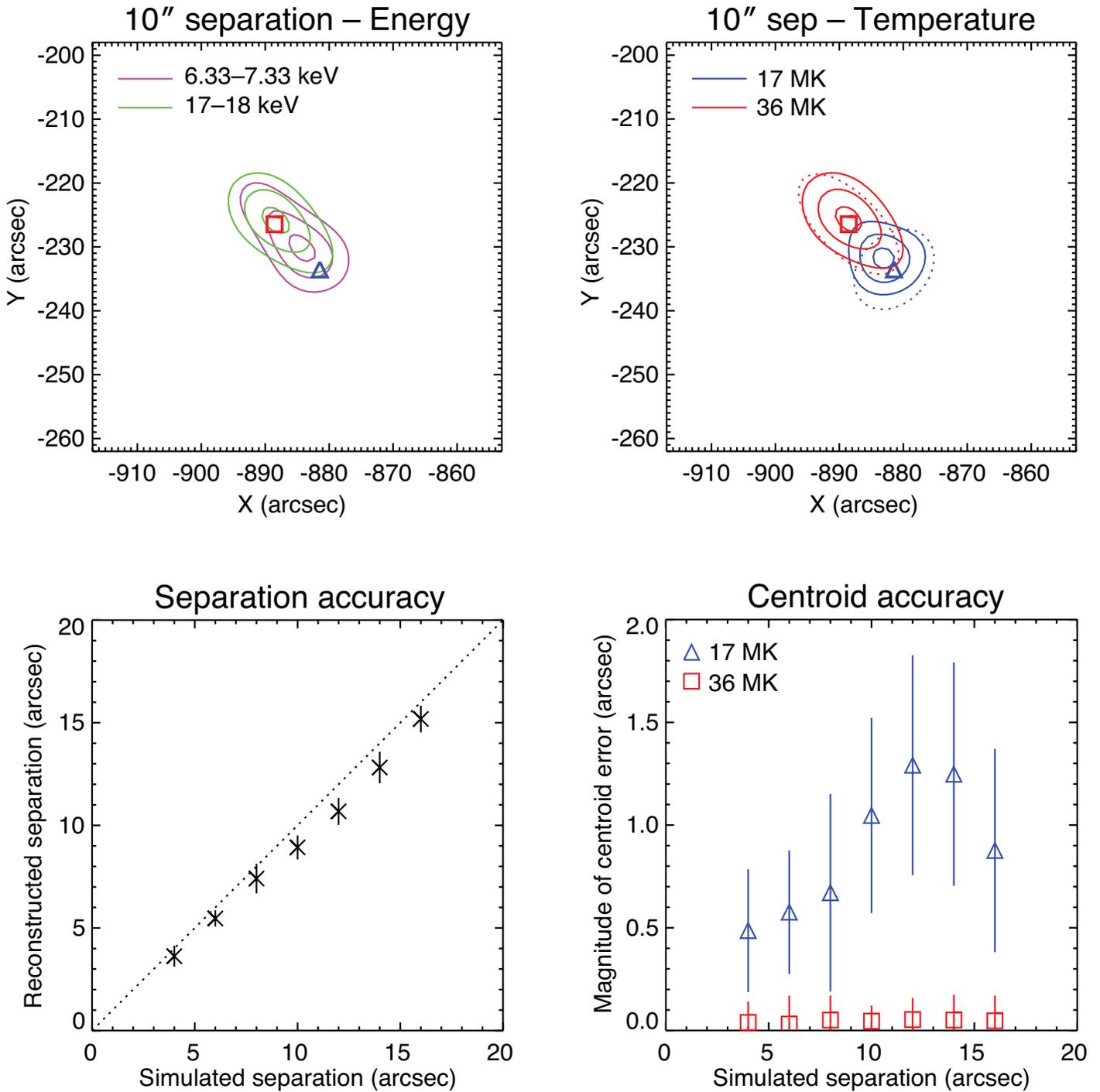

**Figure 5** – *Upper*: An example Monte Carlo simulation of two elliptical Gaussian isothermal sources, in two energy bands (*left*) and after reconstruction of the sources (*right*), with contours at the same levels as in prior figures. The true centroids of the super-hot and cooler sources are represented by *square* and *triangle* symbols, respectively, and the 50% levels of the true sources are overplotted as *dotted* lines. *Lower*: For a range of true source separations, the separation between the reconstructed sources (*left*) and the distances between the reconstructed and true centroids for each source (*right*), with each point representing the mean and standard deviation of all simulations at that source separation.